\documentclass[letterpaper, 10pt, conference]{ieeeconf}
\IEEEoverridecommandlockouts
\usepackage{amsmath,amssymb,url}
\usepackage{graphicx}
\usepackage{color}
\usepackage{siunitx}
\usepackage{algorithm}
\usepackage{algorithmic}
\usepackage[hidelinks]{hyperref}
\usepackage[capitalize,noabbrev]{cleveref}
\usepackage{booktabs}
\usepackage{tikz,pgffor,subcaption}
\usetikzlibrary{arrows.meta}

\newtheorem{theorem}{Theorem}[section]
\newtheorem{proposition}[theorem]{Proposition}

\newtheorem{definition}[theorem]{Definition}
\newtheorem{assumption}[theorem]{Assumption}
\newtheorem{remark}[theorem]{Remark}

\newcommand{\R}{\mathbb{R}}

\newcommand{\E}{\mathbb{E}}

\title{
\LARGE \bf
Max-Entropy Moment Filtering for Stochastic Hybrid Systems
}

\author{Kaito Iwasaki, Tejaswi K. C., Anthony Bloch, Maani Ghaffari, and Taeyoung Lee \vspace{-0.1in}
    \thanks{Kaito Iwasaki, Anthony Bloch, Maani Ghaffari are with the University of Michigan, Ann Arbor MI 48109, USA. {\tt \{kaitoi, abloch, maanigj\} @umich.edu}}%
    \thanks{Tejaswi K. C. and Taeyoung Lee are with The George Washington University, Washington DC 20052, USA. {\tt kctejaswi999@gmail.com, tylee@gwu.edu}}%
}

\graphicspath{{./figures/}}

\begin{document}
	\allowdisplaybreaks
	\maketitle \thispagestyle{empty} \pagestyle{empty}
    
    \begin{abstract}
        Stochastic hybrid systems combine continuous-time stochastic dynamics with discrete reset events, producing intrinsically non-Gaussian and often multimodal uncertainty. A consistent propagation law must also account for boundary-induced probability flux across guard sets, making direct density propagation through hybrid Fokker-Planck equations expensive. We develop a hybrid extension of the Max-Entropy Moment Kalman Filter (MEM-KF) that performs filtering from partial statistical information by propagating a finite collection of moments through stochastic hybrid dynamics and reconstructing beliefs using moment-constrained maximum-entropy distributions. The key step is a moment propagation rule derived from Dynkin's formula with a jump-sum, in which reset effects appear as a boundary-flux correction over the guard set. This yields tractable moment dynamics without solving the underlying hybrid PDE. In a stochastic bouncing-ball example, the proposed method captures reset-induced non-Gaussianity through corrected moment equations while retaining the MEM-KF's optimization-based maximum-entropy representation.
    \end{abstract}

\section{Introduction}

Stochastic hybrid systems model dynamical systems that evolve through continuous-time stochastic dynamics combined with discrete jump events. Such systems arise naturally in a wide range of applications, including mechanical systems with impacts, robotic systems with contact and mode switching, and biological systems with event-driven transitions such as gene regulation or cell-state switching. The hybrid interaction between continuous evolution and discrete resets enables rich modeling expressiveness, but also introduces fundamental challenges for uncertainty representation and propagation.

In this paper, we study the propagation and inference of uncertainty in stochastic hybrid systems from partial statistical information. Two challenges are central: reset events readily induce non-Gaussian, often multi-modal uncertainty, and a propagation law must account for probability mass leaving and re-entering the continuous state space through reset maps.

A direct density-level treatment is expensive in the hybrid setting. The underlying hybrid Fokker--Planck equation couples interior PDE dynamics with nontrivial boundary conditions induced by resets, requiring specialized discretization schemes or spectral methods \cite{WANG2020108989}. Related Frobenius--Perron operator-theoretic approaches have also been developed for deterministic hybrid systems, where the transfer of density across guards and reset maps is characterized precisely through a hybrid Jacobian \cite{Oprea2024}. However, these approaches remain computationally demanding and are not naturally tailored to filtering from partial statistical information.

\begin{figure}[t]
    \centering
    \resizebox{1.29\linewidth}{!}
    {\input{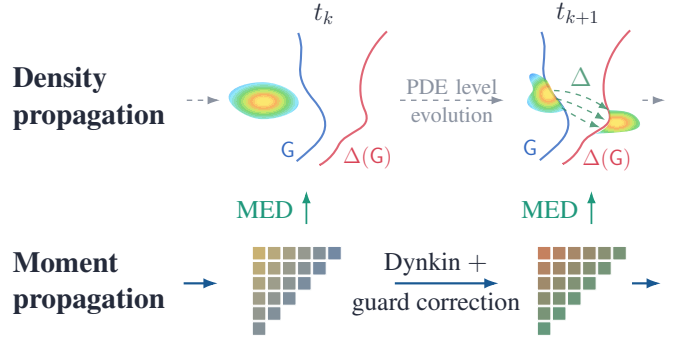}}
    \vspace{0.3em}
    \caption{From density evolution to moment-based uncertainty propagation and reconstruction in SHSs.}
    \label{fig:first-page-schematic}
\end{figure}

To address these challenges, we adopt a weak, expectation-based formulation of uncertainty propagation for stochastic hybrid systems. Rather than evolving full probability densities or individual sample paths, we characterize the stochastic hybrid process through the evolution of expected observables. This leads to an operator-theoretic description via the stochastic Koopman operator, whose infinitesimal generator and associated Dynkin’s formula provide a unified account of both continuous stochastic dynamics and discrete reset events. In particular we derive a hybrid form of Dynkin’s formula in which reset effects appear as boundary flux contributions across the guard set, yielding a consistent evolution law for observables in the presence of jumps.

In practice, only finitely many moments can be propagated, so the underlying distribution is not uniquely determined. We resolve this closure ambiguity with a moment-constrained maximum entropy principle, yielding a finite-dimensional convex reconstruction of uncertainty.

\section{Background}

\subsection{Stochastic Hybrid Systems}\label{sec:SHS}


We adopt the general stochastic hybrid systems viewpoint and focus on the practically important special case of a single discrete mode. This already covers impact-driven mechanical systems, where hybrid behavior arises from guard-triggered state resets rather than switching among several operating modes. The bouncing-ball example considered later is of this type; extension to multiple modes is straightforward, and the results readily generalize to that setting, at the cost of additional notational complexity \cite{KC2025280}.

\begin{definition}[Stochastic hybrid systems]\label{def:SHS}
The stochastic hybrid system is a tuple $\mathcal{H} = (\mathsf{H}, \mathcal{F}, \mu, X, h, \mathsf{G}, K)$ described as follows:
\begin{itemize}
    \item Let $\mathsf{X} \subset \mathbb{R}^n$ be an open domain representing the continuous state space in the single-mode setting. The corresponding hybrid state space is $\mathsf{H} = \mathsf{X} \times \{1\}$, which we identify with $\mathsf{X}$ whenever convenient.
    \item We equip $\mathsf{H}$ with the $\sigma$-algebra $\mathcal{F}$.
    \item The reference measure is $\mu: \mathcal{F} \to \mathbb{R}_{\ge 0}$ on $(\mathsf{H}, \mathcal{F})$.
    \item The continuous state evolves according to the stochastic differential equation
    \begin{equation} \label{eq:shs_cont}
        d x=X(x)\, d t+h(x)\, d W,
    \end{equation}
    where $X(\cdot): \mathsf{X} \to \mathbb{R}^n$ is a drift vector field, $h(\cdot) \in \mathbb{R}^{n \times n_w}$ is the diffusion matrix, and $W$ is an $n_w$-dimensional standard Wiener process.
    \item Let $\mathsf{G} \subset \mathsf{X}$ be a measurable guard set. A discrete transition is triggered when the continuous trajectory reaches $\mathsf{G}$.
    \item During each discrete transition, the hybrid state jumps according to the Dirac reset kernel $K: \mathsf{G} \times \mathcal{F} \to \mathbb{R}_{\ge 0}$
    \begin{equation} \label{eq:shs_disc}
        K(x^{-}, H^+)=\delta_{\Delta(x^-)}(H^+),
    \end{equation}
    where $\Delta: \mathsf{G} \to \mathsf{H}$ is a reset map, so that
    \begin{equation}\label{eq:shs_jump}
    x^{+}=\Delta(x^{-}), \quad x^- \in \mathsf{G}.
    \end{equation}
    Here, $K(x^{-}, H^{+})$ denotes the probability assigned to the measurable set of post-jump states $H^{+} \in \mathcal{F}$, conditioned on the pre-jump state $x^{-}$.
\end{itemize}
We call $(X_t)_{t\ge0}$ a (stochastic) hybrid process if it evolves according to the continuous dynamics and reset rules specified above. The hybrid nature arises from the interplay between the continuous stochastic flow~\eqref{eq:shs_cont} and the discrete transitions~\eqref{eq:shs_disc}-\eqref{eq:shs_jump}. Such a formulation provides a unified framework for modeling systems that combine continuous stochastic dynamics with event-driven state changes. 
\end{definition}

The preceding definitions provide a precise trajectory-level description of stochastic hybrid systems. However, learning and analysis directly at the level of sample paths is challenging due to its noise and discontinuities. Thus, we shift attention from individual trajectories to the evolution of \emph{observables} and their expectations. This perspective leads us to the stochastic Koopman operator which encodes the dynamics of nonlinear stochastic systems via infinite-dimensional linear operators. In what follows we specialize the stochastic hybrid system above to a setting with deterministic boundary resets, and derive the associated infinitesimal generator and distributional evolution.

\subsection{Stochastic Koopman Operator}
Let $X_t$ be a time-homogeneous hybrid process associated with the single-mode system. In the stochastic setting, the Koopman operator is generalized as follows:
\begin{definition}[Stochastic Koopman operator]
    The stochastic Koopman operator $\mathcal{K}_t$ : $L^\infty(\mathsf{H}) \to L^\infty(\mathsf{H})$ for $t>0$ is defined as
    \begin{equation}
    \mathcal{K}_t f(x)=\mathbb{E}[f(X_t) \mid X_0=x], \quad f \in L^\infty(\mathsf{H})
    \end{equation}
    The family $\{\mathcal{K}_t\}_{t\ge0}$ forms a Markov semigroup,
    \begin{equation}
    \mathcal{K}_{t+s}=\mathcal{K}_t\mathcal{K}_s, \quad \mathcal{K}_0=\operatorname{Id},
    \end{equation}
    reflecting the time-homogeneous and Markovian nature of the hybrid process. Between discrete transitions, this semigroup corresponds to the stochastic flow generated by the interior SDE, while resets act through instantaneous composition with the reset map $\Delta$.
\end{definition}
The infinitesimal behavior of the Koopman semigroup leads to the notion of the infinitesimal generator, which links trajectory-level dynamics to distributional evolution.
\subsection{Infinitesimal Generator}
Let $\{\mathcal{K}_t\}_{t\ge0}$ denote the stochastic Koopman semigroup associated with the single-mode hybrid process $X_t$.
\begin{definition}[Infinitesimal generator]\label{def:generator}
The infinitesimal generator $\mathcal{A}$ of the stochastic Koopman semigroup is defined, for functions $f:\mathsf{H}\to\mathbb{R}$ in its domain, by
\begin{equation}
    (\mathcal{A}f)(x):=\lim_{t\downarrow 0}\frac{\mathcal{K}_t f(x) - f(x)}{t},
\end{equation}
whenever the limit exists.
\end{definition}
For the stochastic hybrid system of Definition \ref{def:SHS}, the generator admits an explicit form in the interior of each mode.
\begin{proposition}\label{prop:generator}
Define
\begin{equation}
H(x) := \tfrac12 h(x)h(x)^\top .
\end{equation}
Then, for $f\in C^2(\mathsf H)$ and $x\in \operatorname{int}(\mathsf H)$,
\begin{equation}\label{eq:interior-generator}
(\mathcal{A}f)(x) = \nabla f(x)\cdot X(x) + \operatorname{Tr}(H(x)\nabla^2 f(x)).
\end{equation}
\end{proposition}
This is the standard infinitesimal generator of the interior diffusion and follows directly from It\^o's formula.
The expression \eqref{eq:interior-generator} describes the infinitesimal evolution of observables between discrete transitions. Contributions due to resets at the guard set $\mathsf{G}$ appear as jump terms in the global evolution and will be accounted for via Dynkin’s formula.
\subsection{Dynkin's Formula}
Dynkin’s formula provides a fundamental relationship between the infinitesimal generator of a Markov process and the global evolution of observables along its sample paths. For stochastic hybrid systems, the formula must account for both continuous evolution and discrete reset events. 
\begin{theorem}[Dynkin's formula with jump-sum]\label{thm:dynkin}
Let $X_t$ be a stochastic hybrid process associated with Definition~\ref{def:SHS} with the single-mode assumption, and let $f\in C^2(\mathsf{H})$ with compact support in each mode. Then, for all $t\ge0$,
\begin{equation}\label{eq:dynkin}
\begin{aligned}
&\mathbb{E}[f(X_t)] - f(X_0)\\
&=\mathbb{E}\big[\int_0^t (\mathcal{A}f)(X_s)ds\big] + \mathbb{E}\big[\sum_{0<s\le t}(f(X_s)-f(X_{s^-}))\big],
\end{aligned}
\end{equation}
where the sum is taken over jump times $s$ at which $X_{s^-}\in\mathsf{G}$ and $X_s = \Delta(X_{s^-})$.
\end{theorem}
This is the natural Dynkin identity for the present reset setting. Compare the standard jump-process formulations in \cite{Applebaum2009}.
Note that the first term on the right-hand side of \eqref{eq:dynkin} accounts for the continuous-time evolution of the observable between reset events through the infinitesimal generator $\mathcal{A}$. The second term captures the contribution of discrete reset events at the guard set. The next proposition interprets the jump term in \eqref{eq:dynkin} as a boundary flux across the guard set. This provides a probabilistic description of reset events geometrically, which is crucial to deriving the propagation of observables.
\begin{proposition}\label{prop:jump-flux}
Let $p(t,x)$ denote the probability density of $X_t$ with respect to the reference measure $\mu$, and let
\begin{equation}
J(t,x) := X(x)p(t,x) - \nabla(H(x)p(t,x)) \label{eqn:J}
\end{equation}
denote the associated probability current in the interior. Then,
\begin{equation}\label{eq:jump-flux}
\begin{aligned}
&\mathbb{E}\left[\sum_{0<s\le t}
(f(X_s)-f(X_{s^-}))\right]\\
&= \int_0^t \int_{\mathsf{G}}(f(\Delta(x)) - f(x))(J(\tau,x)\cdot n(x))_+dS(x)d\tau,
\end{aligned}
\end{equation}
where $n(x)$ denotes the outward unit normal on $\mathsf{G}$, $(\cdot)_+ := \max\{\cdot,0\}$, and $dS$ is the surface measure on $\mathsf{G}$. 
\end{proposition}
This gives the jump contribution a boundary-flux representation consistent with the weak hybrid Fokker--Planck viewpoint in \cite{BECT2010357}.
This allows us to describe the time evolution of expected observables along the stochastic hybrid process. Differentiating \eqref{eq:dynkin} and using Proposition \ref{prop:jump-flux}, we obtain the instantaneous evolution
\begin{equation}\label{eq:dynkin-instant}
\begin{aligned}
    &\frac{d}{dt}\mathbb{E}[f(X_t)]=\mathbb{E}[(\mathcal{A}f)(X_t)]\\
    &+\int_{\mathsf{G}}(f(\Delta(x)) - f(x))(J(t,x)\cdot n(x))_+dS(x).
\end{aligned}
\end{equation}
\begin{remark}[Hybrid Fokker-Planck equation]\label{rem:HFP}
Dynkin’s formula \eqref{eq:dynkin-instant} admits a dual interpretation in terms of the evolution of probability densities. Indeed, since
\[
\mathbb{E}[f(X_t)] = \int_{\mathsf{H}} f(x)p(t,x)d\mu(x),
\]
equation \eqref{eq:dynkin-instant} can be viewed as the weak form of a hybrid Fokker-Planck equation governing the density $p(t,x)$. In the interior of each mode, the density satisfies the classical Fokker-Planck equation
\[
\partial_t p(t,x) + \nabla\cdot J(t,x) = 0,
\quad x\in \operatorname{int}(\mathsf{X}),
\]
where $J(t,x)$ defined by \eqref{eqn:J} is the probability current.
At the guard set $\mathsf{G}$, probability mass exits the domain according to the outgoing flux $(J\cdot n)_+$ and is instantaneously reinserted at the reset image $\Delta(\mathsf{G})$. This expresses conservation of probability across hybrid reset events. For more general stochastic hybrid systems, weak formulations of the Fokker--Planck--Kolmogorov equation and sufficient conditions for existence of weak solutions have been studied in \cite{BECT2010357}.
\end{remark}
\section{Problem Formulation}

\subsection{Motivation: state estimation in SHSs}
We are interested in estimating the evolving state of a stochastic hybrid system from noisy measurements, i.e., computing the posterior belief
\begin{equation}
\pi_t(A) = \Pr(X_t \in A \mid \mathcal Y_t), \quad A \in \mathcal{F},
\end{equation}
where $\pi_t$ is a probability measure on the hybrid state space $\mathsf H$ and $\mathcal Y_t := \sigma(\{y_k : t_k \le t\})$ denotes the information generated by all measurements $\{y_k\}_{k=1}^N$ available up to time $t$. We assume a conditional likelihood $p(y_k \mid X_{t_k})$. Exact Bayesian filtering is generally intractable because it requires propagating full densities through coupled diffusion and reset-induced dynamics.
\subsection{Core problem: moment propagation and belief reconstruction}
Rather than directly propagating full probability distributions, we focus on the propagation of \emph{partial statistical information} associated with a stochastic hybrid process. Specifically, given a collection of observables $\{\phi_i\}_{i=1}^M$ on the hybrid state space $\mathsf H$, we consider the time evolution of their expectations
\begin{equation}
m_i(t) := \E[\phi_i(X_t)], \quad t \ge 0,
\end{equation}
where the expectation is taken with respect to the law of the hybrid process $X_t$. We refer to $m_i(t)$ as \emph{(generalized) moments} and collect them into the moment vector $m(t) := (m_1(t),\dots,m_M(t))$.\vspace{0.5em}

\noindent\textbf{Objective.}\quad
The primary objective of this paper is twofold:
\begin{enumerate}
    \item to derive a tractable evolution law for the moment vector $m(t)$ associated with a stochastic hybrid process $X_t$ accounting for both continuous-time stochastic dynamics and reset events;
    \item to reconstruct, at arbitrary times $t$, an approximate belief $\hat\pi_t$ over the hybrid state that is consistent with the available moment information.
\end{enumerate}
Crucially, this problem is posed independently of any measurement model and concerns uncertainty propagation intrinsic to the hybrid dynamics themselves.

\subsection{Relation to filtering}
When noisy measurements are available at discrete (and possibly irregular) times $\{t_k\}_{k\ge0}$, the moment propagation and belief reconstruction machinery developed in this paper can be embedded within a continuous-discrete Bayesian filtering framework. Let $\pi_{t_{k-1}}(\cdot)=\Pr(X_{t_{k-1}}\in\cdot\mid \mathcal Y_{t_{k-1}})$ denote the posterior belief at time $t_{k-1}$. Between measurements $y_{k-1}$ and $y_k$, the belief is propagated forward in time according to the hybrid process:
\begin{equation}\label{eq:prediction}
\pi_{t_k}^-(A) = \Pr(X_{t_k}\in A \mid \mathcal Y_{t_{k-1}}).
\end{equation}
After the new measurement $y_k$ at time $t_k$, Bayes' rule yields the update
\begin{equation}\label{eq:update}
\pi_{t_k}(dx)\propto p(y_k\mid x)\,\pi_{t_k}^-(dx).
\end{equation}
In our framework, the prediction step \eqref{eq:prediction} is performed implicitly via continuous-time propagation of a finite moment vector, and the update \eqref{eq:update} is incorporated at the moment level followed by belief reconstruction.

\subsection{Uncertainty Propagation via Moment Equations}\label{sec:moment propagation}

In what follows, we work exclusively with the weak formulation \eqref{eq:dynkin-instant}, which suffices for deriving moment equations and learning dynamics from data, without explicitly solving the underlying hybrid PDE.

The hybrid Fokker-Planck equation provides a complete description of the distributional evolution of the stochastic hybrid process. However, directly working with probability densities is often impractical, especially in high dimensions or when only partial statistical information is available. Instead, we characterize the evolution of the process through a finite collection of moments. First, we derive closed-form evolution equations for moments of the hybrid process using Dynkin’s formula. 

For polynomial observables in the continuous state, writing $x\in\mathsf{H} \cong \mathsf{X}\subset \mathbb{R}^n$ with the single-mode system as before, let $\alpha\in\mathbb{N}^n$ be a multi-index and define the monomial observable
\begin{equation}
\phi_\alpha(x) := x^\alpha.
\end{equation}
We denote its corresponding moment of the hybrid process $X_t$ as
\begin{equation}
m_\alpha(t) := \mathbb{E}[\phi_\alpha(X_t)].
\end{equation}
Applying Dynkin’s formula \eqref{eq:dynkin-instant} with $f=\phi_\alpha$ yields an evolution equation for the moment $m_\alpha(t)$:
\begin{equation}\label{eq:moment-dynkin}
\begin{aligned}
&\frac{d}{dt} m_\alpha(t) = \mathbb{E}[(\mathcal{A}\phi_\alpha)(X_t)] \\
&\quad + \int_{\mathsf{G}}(\phi_\alpha(\Delta(x)) - \phi_\alpha(x)) (J(t,x)\cdot n(x))_+dS(x).
\end{aligned}
\end{equation}
For polynomial observables, if the stochastic system is a polynomial system (i.e., $X$ and $h$ are polynomials), the generator term $\mathcal{A}\phi_\alpha$ is again a polynomial function of the state. Consequently, the right-hand side of \eqref{eq:moment-dynkin} depends on moments while the reset term couples moments before and after impact through the reset map $\Delta$. 

\section{Moment-constrained Maximum Entropy Distribution}\label{sec:MED}

In the stochastic hybrid setting considered here, the evolution of finitely many moments can be characterized exactly via Dynkin’s formula \eqref{eq:moment-dynkin}. While an infinite sequence of moments may uniquely determine a distribution under suitable regularity conditions, such information is neither available nor tractable in practice. As a result, we are typically limited to finite moment information, which does not uniquely determine the underlying state distribution. This non-uniqueness leads to an inherent closure ambiguity. To resolve this ambiguity in a principled manner, we employ the \emph{Moment-Constrained Maximum Entropy Principle}, which selects the distribution with the maximum entropy among all distributions consistent with the prescribed moment constraints \cite{MeadPapanicolaou1984}.

To obtain a finite-dimensional approximation, we fix a maximal total degree $r\ge1$ and define a set $\mathbb{N}_r^n = \{\alpha \in \mathbb{N}^n: |\alpha|= \sum_i \alpha_i \le r\}$.
We track only finitely many moments
\begin{equation}
m_\alpha(t) := \mathbb{E}[\phi_\alpha(X_t)],
\quad \alpha\in \mathbb{N}^n_r.
\end{equation}
We collect these moments into the truncated moment vector
\begin{equation}
m(t) := (m_\alpha(t))_{\alpha \in \mathbb{N}^n_r}.
\end{equation}
Let $\mu$ denote the reference measure as in Definition~\ref{def:SHS}. We denote the set $\mathcal{P}_r(m)$ of admissible probability densities
\begin{equation}
\mathcal{P}_r(m) := \{p\ge0 :\langle 1,p\rangle_\mu = 1,\langle x^\alpha,p\rangle_\mu = m_\alpha,
\alpha\in \mathbb{N}_r^n\},
\end{equation}
where $\langle f, p\rangle_\mu:=\int_{\mathsf{H}} f(x) p(x) d \mu(x)$. For strict feasibility and interior realizability of the subsequent problem, we make the following assumption.

\begin{assumption}\label{assump:strict feasibility/interior realizability}
Assume $\mathcal{P}_r(m) \neq \emptyset$ and that there exists $p_0 \in \mathcal{P}_r(m)$ such that $p_0> 0$ $\mu$-a.e. on $\mathsf{H}$ and $\int_\mathsf{H}|x^\alpha|p_0 d\mu < \infty$ for all $\alpha \in \mathbb{N}^n_r$.
\end{assumption}

The MED can be obtained by maximizing the entropy functional over $\mathcal{P}_r(m)$:
\begin{equation}
p_r^*:= \arg\max_{p\in\mathcal{P}_r(m)}
\left(-\int_{\mathsf H} p(x)\log p(x)d\mu(x)\right).
\end{equation}
By the first-order optimality condition, the solution $p_r^*$, when it exists, belongs to a finite-dimensional exponential family,
\begin{equation}\label{eq:MEM-density}
p_r^*(x;\lambda)= \frac{1}{Z(\lambda)}\exp\Big(-\sum_{\alpha}\lambda_\alpha x^\alpha\Big),
\end{equation}
where $\lambda=(\lambda_\alpha)$ are Lagrange multipliers associated with the moment constraints, and the partition function $Z(\lambda)$ is defined as
\begin{equation}
Z(\lambda):= \int_{\mathsf X}\exp\Big(-\sum_{\alpha }\lambda_\alpha x^\alpha\Big)d\mu(x).
\end{equation}
To compute the maximum-entropy density in practice, it is convenient to work with the associated dual objective called the \emph{thermal dynamics potential}\footnote{$\Gamma(\lambda)$ is obtained from a modified Legendre–Fenchel dual of the entropy functional, in which the normalization constraint is absorbed into the log-partition function $Z(\lambda)$.}
\begin{equation}\label{eq:thermal-potential}
\Gamma(\lambda):= \log Z(\lambda) + \sum_{\alpha}\lambda_\alpha\,m_\alpha,
\end{equation}
defined on the open domain $\operatorname{dom}(Z):=\{\lambda:\ Z(\lambda)<\infty\}$. 

\noindent Differentiating $\Gamma$ yields
\begin{equation}
\frac{\partial \Gamma(\lambda)}{\partial \lambda_\beta}
= -\int_{\mathsf{X}} x^\beta p_r^*(x;\lambda) d\mu(x) + m_\beta,
\end{equation}
so that any critical point of $\Gamma$ satisfies the moment-matching conditions. Furthermore, the Hessian of $\Gamma$ is given by
\begin{equation}
\begin{aligned}
&\frac{\partial^2 \Gamma(\lambda)}{\partial \lambda_\beta\partial \lambda_\gamma} =\int_{\mathsf{H}} x^\beta x^\gamma p_r^*(x;\lambda) d \mu(x)\\
&-
\left(\int_{\mathsf{H}} x^\beta p_r^*(x;\lambda) d\mu(x)\right)
\left(\int_{\mathsf{H}} x^\gamma p_r^*(x;\lambda) d\mu(x)\right),
\end{aligned}
\end{equation}
which is the covariance matrix of the sufficient statistics $\{x^\alpha\}$. Hence $\Gamma$ is convex on $\operatorname{dom}(Z)$. Under Assumption \ref{assump:strict feasibility/interior realizability}, the covariance matrix is positive definite, implying the strict convexity of $\Gamma$. We therefore define
\begin{equation}\label{eq:MED parameters}
    \lambda^* = \arg\min_{\lambda \in \operatorname{dom}(Z)} \Gamma(\lambda).
\end{equation}
The corresponding density $p_r^*(x;\lambda^*)$ provides the moment-constrained maximum-entropy approximation consistent with the prescribed moment vector $m$.

\section{Algorithm}\label{sec:algorithm}
Algorithm~\ref{alg:MEMKF_for_SHS} summarizes the proposed max-entropy moment filter for stochastic hybrid systems.
\subsection{Initialization}
Given a truncation order $r$ and initial moments $m(t_0)=\{m_\alpha(t_0)\}_{|\alpha| \leq r}$, the initial belief is represented by the MED \eqref{eq:MEM-density} with parameters obtained from \eqref{eq:MED parameters}.
\subsection{Prediction Step}
Between measurement times, $m(t)$ evolves according to \eqref{eq:moment-dynkin}. Numerical integration yields predicted moments $m^-(t+\Delta t)$, from which the predicted MED parameters $\lambda^-(t+\Delta t)$ are recovered via \eqref{eq:MED parameters}.
\subsection{Update Step}\label{subsec:update}
At a measurement time $t_k$, we update the predicted belief using a maximum-entropy model of the measurement residual. We do not assume a known likelihood $p(y\mid x)$. Instead, we assume the measurement function to be polynomial $v = g(y,x)$ and model the residual $v$ by a moment-constrained MED.

Let $\{\psi_\alpha\}_{|\alpha|\le r}$ be a polynomial basis on the residual space and
let $\bar v_{\alpha,k}$ denote prescribed residual moments at time $t_k$.
We define the residual MED
\begin{equation}\label{eq:residual_MED}
p(v;\mu_k)=\frac{1}{Z_v(\mu_k)}\exp\Big(-\sum_{\alpha}\mu_{\alpha,k}\psi_\alpha(v)\Big),
\end{equation}
where $Z_v(\mu_k)$ is the normalization, and $\mu_k$ is obtained from $\bar v_k$ via \eqref{eq:MED parameters}. Substituting $v=g(y,x)$ into \eqref{eq:residual_MED} yields an implicit likelihood
\begin{equation}\label{eq:induced_likelihood}
p(y_k\mid x) \propto \exp\Big(-\sum_{\alpha}\mu_{\alpha,k}\psi_\alpha(g(y_k,x))\Big).
\end{equation}
Let the predicted prior be $p(x;\lambda^-(t_k))$ in \eqref{eq:MEM-density}. Then Bayes' rule implies the posterior is again MED:
\begin{equation}\label{eq:posterior_MED}
\begin{aligned}
    p(x;\lambda^+(t_k)) &\propto p(x;\lambda^-(t_k))p(y_k\mid x),
\end{aligned}
\end{equation}
Finally, the resulting unnormalized density is normalized, yielding the posterior moments
\begin{equation}\label{eq:moment update}
m_\alpha^+(t_k) = \frac{\int x^\alpha p(x;\lambda^-(t_k))\exp(-\sum_{\beta}\nu_{\beta}(y_k)x^\beta)d\mu(x)}{\int p(x;\lambda^-(t_k)) \exp(-\sum_{\beta}\nu_{\beta}(y_k)x^\beta) d\mu(x)},
\end{equation}
and the posterior MED parameters are computed via \eqref{eq:MED parameters}.

\subsection{State Estimation}\label{subsec:state-estimation}
At each time step, the belief is represented by the MED
\[
p(x;\lambda)=\frac{1}{Z(\lambda)}\exp\!\Big(-\sum_{|\alpha|\le r}\lambda_\alpha x^\alpha\Big)
\]
supported on the feasible hybrid domain $\mathsf{H} \subset \mathbb{R}^n$. A point estimate is obtained via maximum a posteriori (MAP) estimation. Since the normalization constant $Z(\lambda)$ does not depend on $x$, this is equivalent to solving the polynomial optimization problem (POP)
\begin{equation}\label{eq:map-pop}
x^*=\arg\max_{x\in\mathsf H} p(x;\lambda)
=\arg\min_{x\in\mathsf H} \sum_{|\alpha|\le r}\lambda_\alpha x^\alpha.
\end{equation}
In practice, this problem can be handled via semidefinite moment relaxations, yielding a semidefinite program \cite{teng2025max}.
\begin{remark}[Semialgebraic representation of $\mathsf{H}$]
    We assume $\mathsf H$ is a basic closed semialgebraic set of the form
\begin{equation}\label{eq:app_semialg}
\mathsf H=\{x\in\R^n:\ g_j(x)\ge 0,\ j=1,\dots,m\},
\end{equation}
where each $g_j$ is a polynomial. In our examples, $\mathsf H$ typically encodes simple bounds (e.g., $x_1\ge 0$ and box constraints for our numerical example) and can always be written in the form \eqref{eq:app_semialg}.
We further assume an \emph{archimedean} condition (e.g., $\mathsf H$ compact or made compact by adding a redundant ball constraint), which guarantees convergence of the relaxation hierarchy \cite{Lasserre2001}. These assumptions are standard in the moment-SDP literature and serve only to justify the polynomial optimization step used for state estimation. Since this is not a main contribution of the present paper, we refer the reader to standard texts, e.g. Lasserre \cite{Lasserre2015}, for the underlying theory.
\end{remark}

\begin{algorithm}[tb]
\caption{Max-Entropy Moment Filtering for SHS}
\label{alg:MEMKF_for_SHS}
\begin{algorithmic}
\STATE \textbf{Require:} Truncation order $r$, time span $[t_0,t_f]$, initial moments $m(t_0)$, SHS model $\mathcal{H}$, measurement model $v = g(y,x)$, measurement schedule $\{t_k\}_{k=1}^N$
\STATE \textbf{Output:} posterior moments $\{m^+(t)\}_{t_0 \le t \le t_f}$, MED parameters $\{\lambda^+(t)\}_{t_0 \le t \le t_f}$
\STATE \textbf{Initialize:} $\lambda^+(t_0)\xleftarrow{\eqref{eq:MED parameters}} m(t_0)$
\WHILE{$t < t_f$}
    \STATE \textbf{Prediction:}
    \STATE \texttt{// Obtain predicted moments}
    \STATE $m^-(t+\Delta t)\xleftarrow{\eqref{eq:moment-dynkin}} (m^+(t),\Delta t)$
    \STATE \texttt{// Reconstruct predicted MED}
    \STATE $\lambda^-(t+\Delta t)\xleftarrow{\eqref{eq:MED parameters}} m^-(t+\Delta t)$
    \STATE \textbf{Measurement update:}
    \IF{$k\le N$ \AND $t+\Delta t =  t_k$}
        \STATE \texttt{// Update posterior moments}
        \STATE $m^+_\alpha(t_k)\xleftarrow{\eqref{eq:moment update}} \lambda^-(t_k)$
        \STATE \texttt{// Reconstruct posterior MED}
        \STATE $\lambda^+(t_k)\xleftarrow{\eqref{eq:MED parameters}} m^+(t_k)$
        \STATE $k \leftarrow k+1$
    \ELSE
        \STATE \texttt{// No measurement}
        \STATE $m^+(t+\Delta t)=m^-(t+\Delta t)$, \quad $\lambda^+(t+\Delta t)= \lambda^-(t+\Delta t)$
    \ENDIF
    \STATE \textbf{State estimation:}
    \STATE \texttt{// Estimate optimal state}
    \STATE $x^* \xleftarrow{\eqref{eq:map-pop}} \min_{x \in \mathsf{H}}-\ln p(x;\lambda^+)$
    \STATE \texttt{// Advance time}
    \STATE $t\leftarrow t+\Delta t$
\ENDWHILE
\end{algorithmic}
\end{algorithm}

\section{Numerical Example}
\subsection{Bouncing Ball}
Let us consider the bouncing ball example which is a single mode stochastic hybrid system. 
The ball evolves continuously under gravity along with noise in the dynamics, while it undergoes instantaneous resets (jumps) when it hits the ground. 
Let the continuous state $x = (x_1, x_2) \in \mathsf{H} = \{x_1 \ge 0, x_2 \in \mathbb{R}\}$ denote height and velocity. The SDE in the interior is
\begin{equation}\label{eq:bb SHS}
    \begin{aligned}
        dx_1 & = x_2 dt,\\
        dx_2 &= (-g - \nu x_2) dt + \sigma_v dW_t,
    \end{aligned}
\end{equation}
where $g > 0$ is gravitational acceleration, $\nu > 0$ is drag-coefficient and $\sigma_v$ controls the noise. The guard (impact surface) is
\begin{equation}
    \mathsf G = \{x_1 = 0, \ x_2 < 0\},
\end{equation}
with a deterministic reset map $\Delta: \mathsf G \to \mathsf H$
\begin{equation}\label{eq:bb impact}
    \Delta(0,x_2) = (0,-cx_2),
\end{equation}
where $0 < c < 1$ is the restitution coefficient. From \eqref{eq:interior-generator}, we derive the infinitesimal generator $\mathcal{A}$ in the interior to be
\begin{equation}\label{eq:bb generator}
    \mathcal{A} f=x_2 \frac{\partial f}{\partial x_1}+\left(-g-\nu x_2\right) \frac{\partial f}{\partial x_2}+\frac{1}{2} \sigma_v^2 \frac{\partial^2 f}{\partial x_2^2}
\end{equation}
\subsubsection{Moment Propagation}
To propagate moments, we need to derive the right-hand-side of equation \eqref{eq:moment-dynkin} specialized to this example. Choose monomial test functions $f(x) = x_1^{\alpha_1}x_2^{\alpha_2}$ and define multi-index moments up to total order $r$
\begin{equation}
    m_\alpha (t) := m_{\alpha_1,\alpha_2}(t) = \mathbb{E}[x_1^{\alpha_1}x_2^{\alpha_2}],
\end{equation}
where $|\alpha| = \alpha_1 + \alpha_2 \le r$. Then, from \eqref{eq:moment-dynkin}, the moment dynamics are
\begin{equation}\label{eq:moment ode bb}
    \dot m_\alpha = \mathbb{E}[\mathcal{A}(x_1^{\alpha_1}x_2^{\alpha_2})] + \Delta_\alpha.
\end{equation}
The generator $\mathcal{A}(x_1^{\alpha_1}x_2^{\alpha_2})$ can be written in a closed form as
\begin{equation}
    \begin{aligned}
        \mathcal{A}(x_1^{\alpha_1} x_2^{\alpha_2})&=\alpha_1 x_1^{\alpha_1-1} x_2^{\alpha_2+1}\\&+\alpha_2(-g-\nu x_2) x_1^{\alpha_1} x_2^{\alpha_2-1}\\&+\frac{1}{2} \alpha_2(\alpha_2-1) \sigma_v^2 x_1^{\alpha_1}x_2^{\alpha_2-2}.
    \end{aligned}
\end{equation}
By taking the expectation, this yields
\begin{equation}
    \begin{aligned}
        \mathbb{E}[\mathcal{A}(x_1^{\alpha_1} x_2^{\alpha_2})] &= \alpha_1 m_{\alpha_1-1, \alpha_2+1} \\&+\alpha_2(-g m_{\alpha_1, \alpha_2-1}-\nu m_{\alpha_1, \alpha_2})\\
        &+\frac{1}{2} \alpha_2(\alpha_2-1) \sigma_v^2 m_{\alpha_1, \alpha_2-2}.
    \end{aligned}
\end{equation}
The impact affects the moments through the velocity reset at the guard $\mathsf G$. A simple observation is that any moment involving height $x_1$ is insensitive to this reset, i.e., if $\alpha_1 \geq 1$, then $x_1^{\alpha_1}=0$ both just before and just after the impact, since the jump occurs at $x_1=0$. Hence $\Delta_\alpha=0$ for all $\alpha_1 \geq 1$. Thus, the only nontrivial correction comes from pure velocity moments $m_{0, \alpha_2}$. Using the reset $x_2^{+}=-c x_2^{-}$, the jump in the test function $f(x)=x_2^{\alpha_2}$ is $f(\Delta(x))-f(x)=((-c)^{\alpha_2}-1) x_2^{\alpha_2}$. Weighting this by the outward probability flux across the guard yields
\begin{equation}\label{eq:boundary flux bb}
\Delta_{0, \alpha_2}=-((-c)^{\alpha_2}-1) \int_{-\infty}^0 x_2^{\alpha_2+1} p(t, 0, x_2) d x_2 .
\end{equation}
Defining the boundary source term $S_{\alpha_2}(t):=-\int_{-\infty}^0 x_2^{\alpha_2+1} p(t, 0, x_2) d x_2$, we obtain the compact form
\begin{equation}
\Delta_\alpha= \begin{cases}0, & \alpha_1 \geq 1, \\((-c)^{\alpha_2}-1) S_{\alpha_2}(t), & \alpha_1=0 .\end{cases}
\end{equation}
Note that at each time step, evaluating the right-hand side of \eqref{eq:moment ode bb} requires the knowledge of the density to compute $S_{\alpha_2}$. Thus, in practice, $\Delta_\alpha$ is approximated using the MED obtained in the previous time step.
\subsubsection{Propagation Steps}
We first study uncertainty propagation without measurements. The initial state distribution is chosen as a Gaussian centered at $(x_1,x_2)=(1.5,0)$ with moderate variance in both position and velocity to model initial uncertainty.  Moments up to total degree $r=4$ are propagated forward in time by numerically integrating the moment ODE \eqref{eq:moment ode bb}. The boundary correction terms $\Delta_\alpha$ are evaluated using a moment-constrained MED reconstructed from the moment vector at the previous time step.

At each time step, a MED of the form \eqref{eq:MEM-density} is fitted to the propagated moments, yielding a continuous approximation of the evolving density on the truncated state space $\mathsf H = [0,3]\times[-6,6]$. This reconstructed density is used both to evaluate the boundary source terms $S_{\alpha_2}(t)$ in \eqref{eq:boundary flux bb} and to visualize the uncertainty evolution.

\begin{figure}[t]
    \centering
    \includegraphics[width=\linewidth]{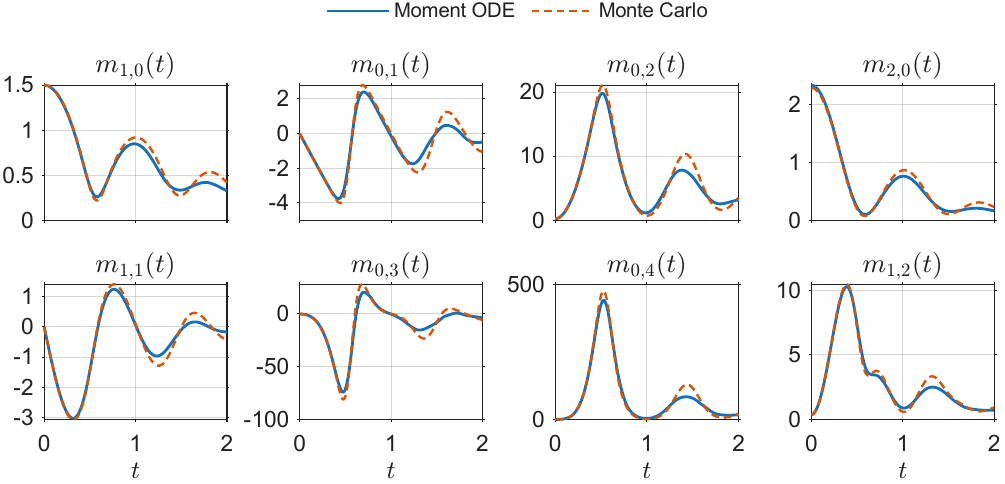}
    \caption{Comparison of moment dynamics for the stochastic bouncing-ball system with Monte Carlo estimates for selected low- and higher-order moments $m_{\alpha}(t)=\mathbb{E}\left[x_1^{\alpha_1} x_2^{\alpha_2}\right]$.}
    \label{fig:bb moments}
\end{figure}

\begin{figure}[t]
    \centering
    \includegraphics[width=0.8\linewidth]{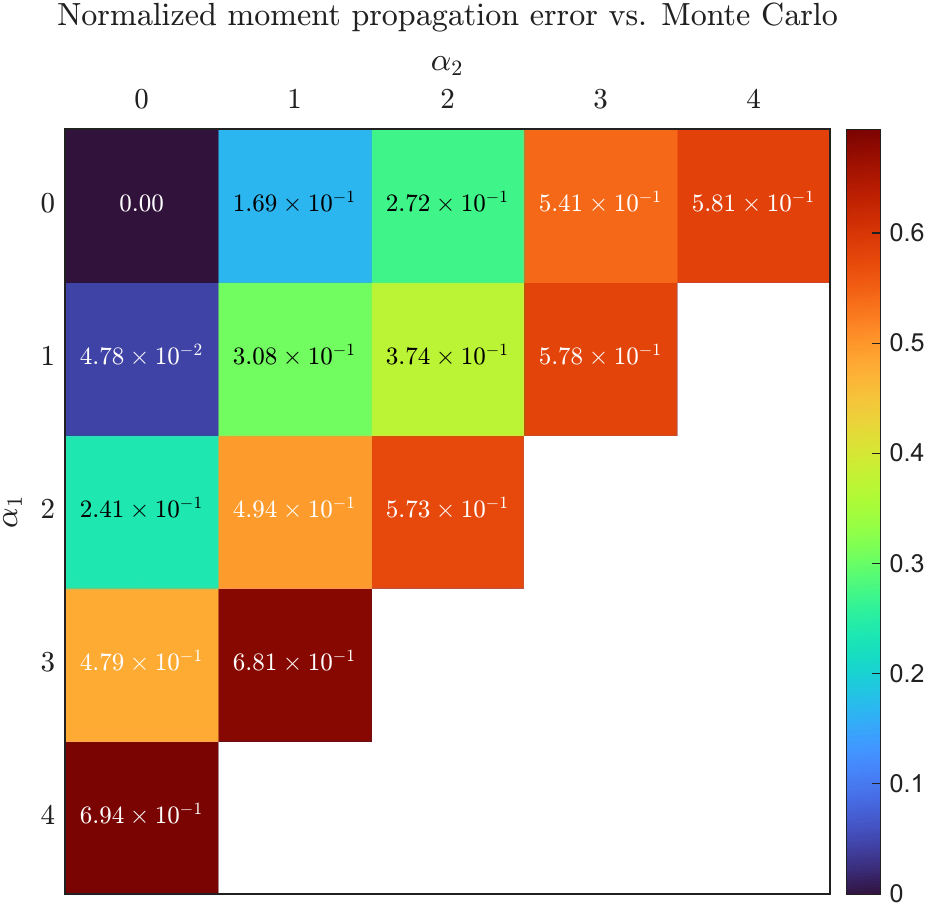}
    \caption{Heat map of normalized rollout errors for propagated moments up to degree $4$ relative to Monte Carlo moments.}
    \label{fig: rmse}
\end{figure}

To assess the accuracy of the moment-based propagation in our bouncing ball stochastic hybrid system, we compute Monte Carlo reference moments by simulating an ensemble of independent sample trajectories of the stochastic hybrid bouncing ball system. For each time $t$, the Monte Carlo estimate of the moment $m_{\alpha}(t) = \mathbb{E}[x_1(t)^{\alpha_1} x_2(t)^{\alpha_2}]$ is obtained by empirical averaging over the ensemble,
\begin{equation}
m^{\mathrm{MC}}_{\alpha}(t) = \frac{1}{N}\sum_{i=1}^{N} x_{1,i}(t)^{\alpha_1} x_{2,i}(t)^{\alpha_2},
\end{equation}
where $\{(x_{1,i}(t),x_{2,i}(t))\}_{i=1}^N$ denote the simulated sample paths. The same initial distribution, noise parameters, and physical constants are used in both the Monte Carlo simulations and the moment ODE propagation. Figure~\ref{fig:bb moments} compares representative propagated moments $m_\alpha(t)$ with Monte Carlo estimates. To provide a more compact summary over all propagated moments up to total degree $4$, Figure~\ref{fig: rmse} shows a heat-map matrix whose $(\alpha_1,\alpha_2)$ entry is a normalized rollout error between the propagated moment and its Monte Carlo reference. More precisely, for each moment we compute a normalized rollout error by taking the root-mean-square of the $(\alpha_1+\alpha_2)$th root of the absolute difference from the Monte Carlo moment, and then dividing by the corresponding root-mean-square Monte Carlo magnitude. This yields a dimensionless quantity that makes errors across different moment orders more comparable.

\begin{figure}[t]
    \centering

    \vspace{-1em}
    \begin{subfigure}{\linewidth}
        \centering
        \vspace{0.3em}
        \caption{Low-order truncation ($r=2$)}
        \vspace{0.3em}
        \includegraphics[width=\linewidth]{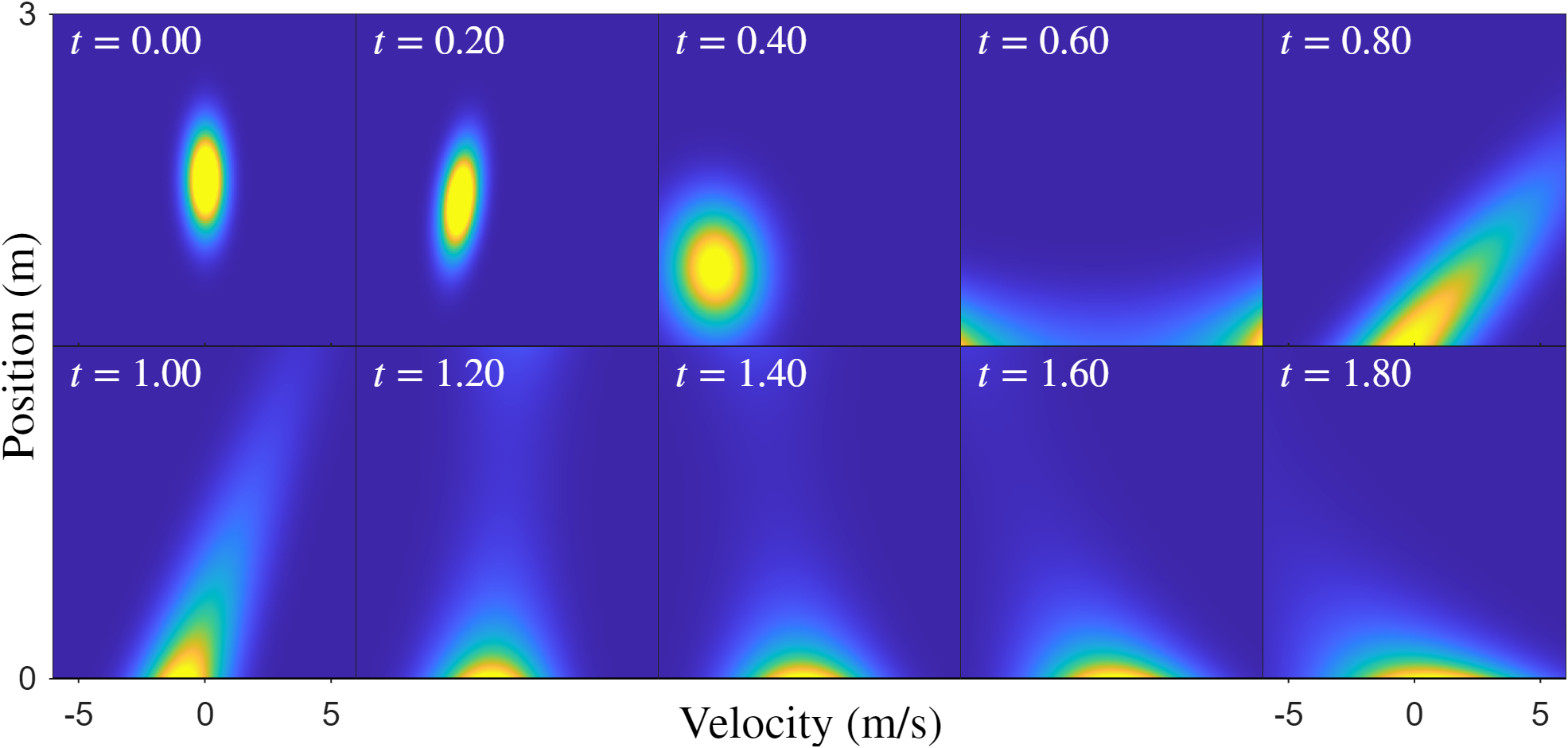}
        \label{fig:bb_density_R2}
    \end{subfigure}

    \vspace{-1.5em}
    
    \begin{subfigure}{\linewidth}
        \centering
        \caption{Higher-order truncation ($r=4$)}
        \vspace{0.3em}
        \includegraphics[width=\linewidth]{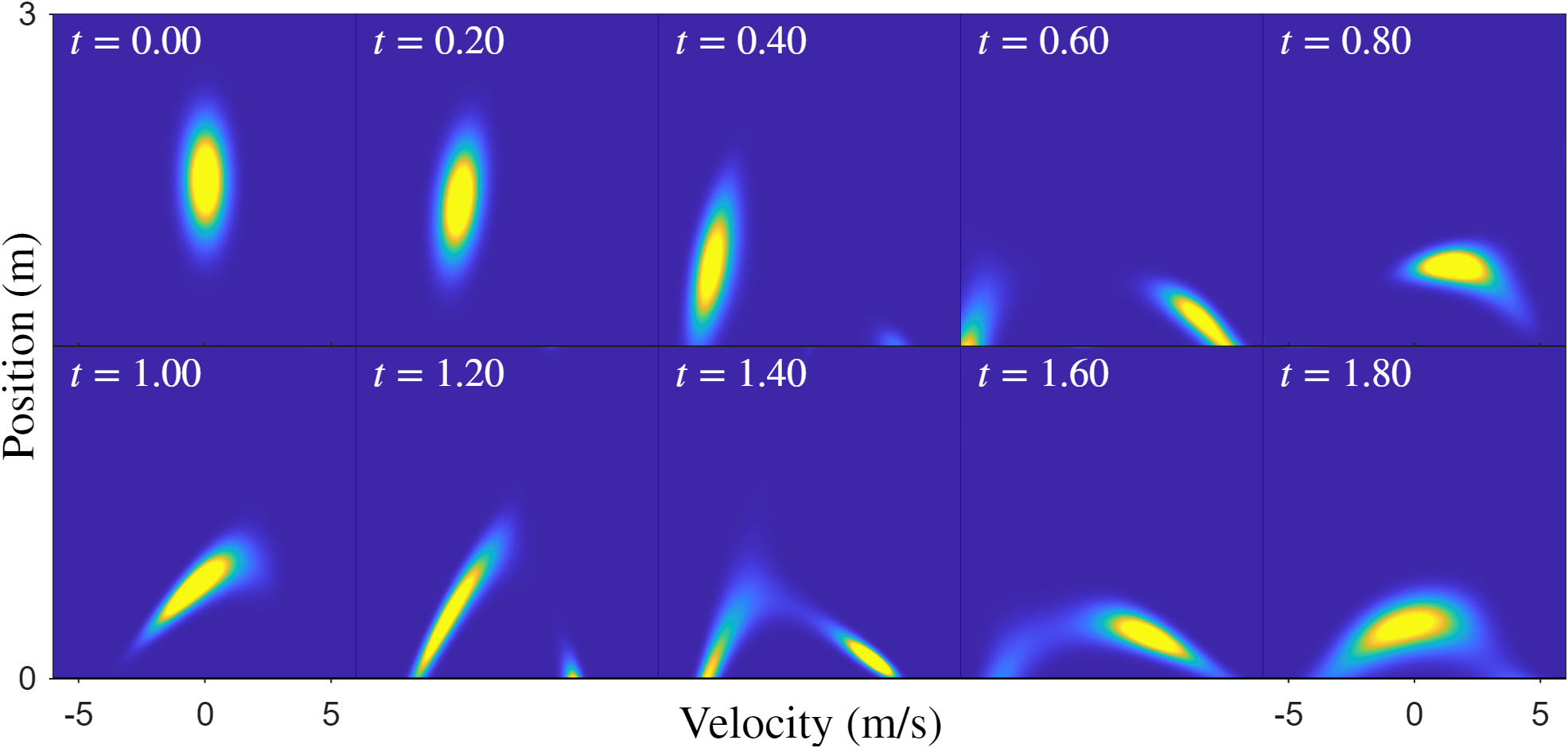}
        \label{fig:bb_density_R4}
    \end{subfigure}

    \vspace{-1.5em}
    
    \begin{subfigure}{\linewidth}
        \centering
        \caption{Monte Carlo reference ($N=2\times 10^5$)}
        \vspace{0.3em}
        \includegraphics[width=\linewidth]{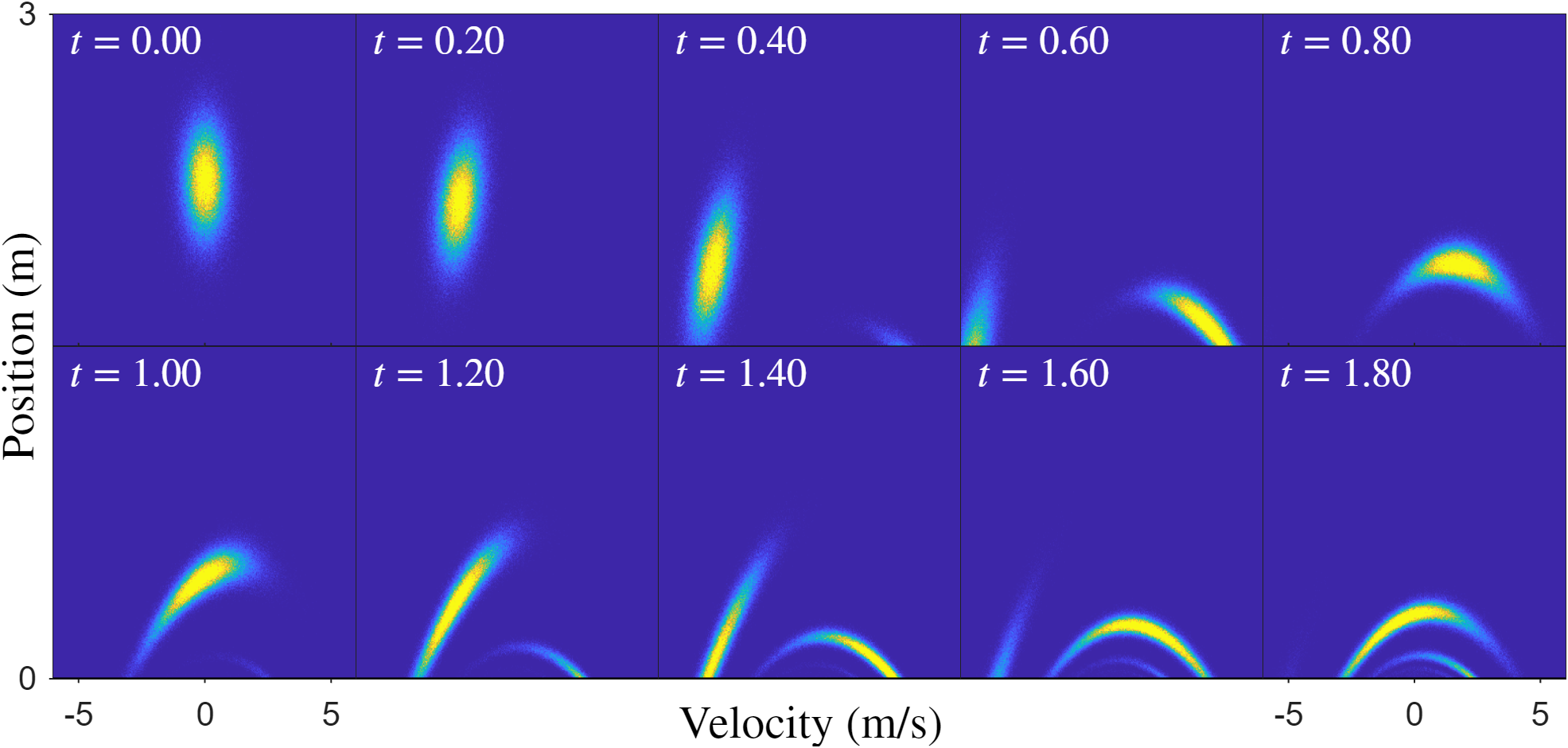}
        \label{fig:bb_density_MC}
    \end{subfigure}
    
    \caption{Propagation-only density evolution for the stochastic bouncing-ball system in the velocity-position plane. Panels (a) and (b) show MED reconstructions from propagated moments for truncation orders $r=2$ and $r=4$, respectively, and panel (c) shows the corresponding Monte Carlo density snapshots. The low-order truncation misses much of the impact-induced deformation, while the higher-order truncation captures the post-impact structure more faithfully and agrees more closely with the Monte Carlo reference.}
    \label{fig:bb propagation only}
\end{figure}

Figure~\ref{fig:bb propagation only} shows snapshots of the corresponding maximum-entropy density reconstructions at selected times. As the ball undergoes repeated impacts, the initially unimodal velocity distribution progressively develops skewness and multimodal structure due to the asymmetric reset $x_2^{+}=-c x_2^{-}$ and the accumulation of probability mass near the guard. These non-Gaussian features are reflected in higher-order moments and are qualitatively captured by the MED reconstructions when the truncation order is set to $r=4$. In contrast, when only second-order moments ($r=2$) are retained, the maximum-entropy reconstruction is necessarily Gaussian and therefore unimodal, restricting the method to a widely used estimation regime in which only the mean and covariance are tracked, as in standard estimators such as the Best Linear Unbiased Estimator (BLUE) \cite{Kay1993}. As a result, non-Gaussian structure induced by impact events cannot be represented in this regime, highlighting the role of higher-order moments in capturing the hybrid nature.

\subsubsection{Measurements and State Estimation}
We now incorporate noisy position measurements to perform state estimation. Position-only measurements are taken at discrete times $\{t_k\}$ and satisfy 
\begin{equation}
    z_k = x_1(t_k) + v_k,
\end{equation}
corrupted by a non-Gaussian noise
\begin{equation}
    v_k = 0.05(2b_k-1) + \epsilon_k,
\end{equation}
with $b_k \sim \mathrm{Bernoulli}(0.5)$ and $\epsilon_k \sim \mathcal{N}(0,0.1^2)$.

\begin{figure}[t!]
    \centering
    {
  \textcolor{blue}{---}\ \textbf{Estimate}
  \quad
  \textcolor{black}{---}\ \textbf{Ground truth}}
    \includegraphics[width=\linewidth]{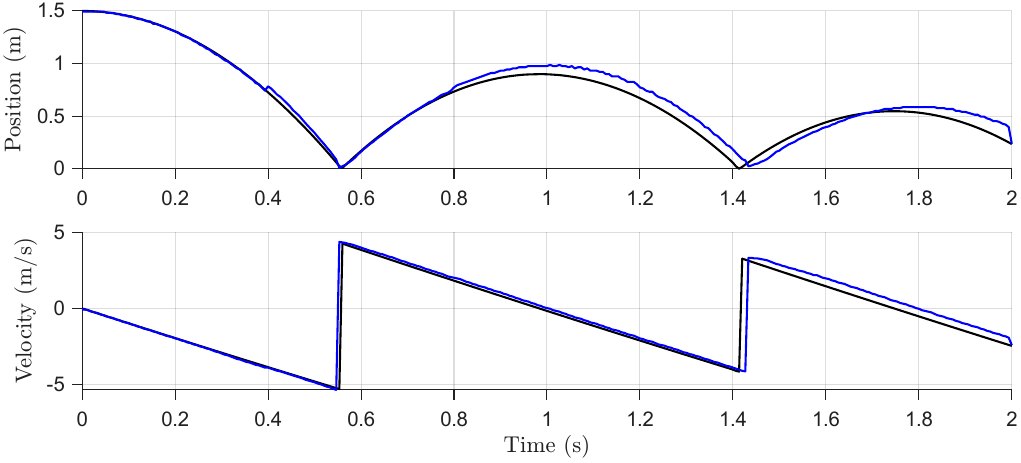}
    \caption{True and estimated trajectories for position and velocity. Despite non-Gaussian uncertainty and discontinuous velocity resets, the MEM-based filter tracks the state accurately.}
    \label{fig:bb trajectories}
\end{figure}

Figure~\ref{fig:bb filtering density} shows snapshots of the reconstructed posterior density at representative times. Figure~\ref{fig:bb trajectories} compares the true state, the estimated state trajectory, and noisy position only measurements over time. Over the full time interval, the filter achieves a rollout RMSE of $6.91\times 10^{-2}$ m in position and $8.62\times 10^{-1}$ m/s in velocity.

\newcommand{\bigbullet}{{\scalebox{1.0}{$\bullet$}}}

\begin{figure}[t!]
    \centering
    {
  \textcolor{red}{\bigbullet}\ \textbf{Estimate}
  \quad
  \textcolor[rgb]{0,0.75,0}{\bigbullet}\ \textbf{Ground truth}}
    \vspace{0.5em}

    \includegraphics[width=\linewidth]{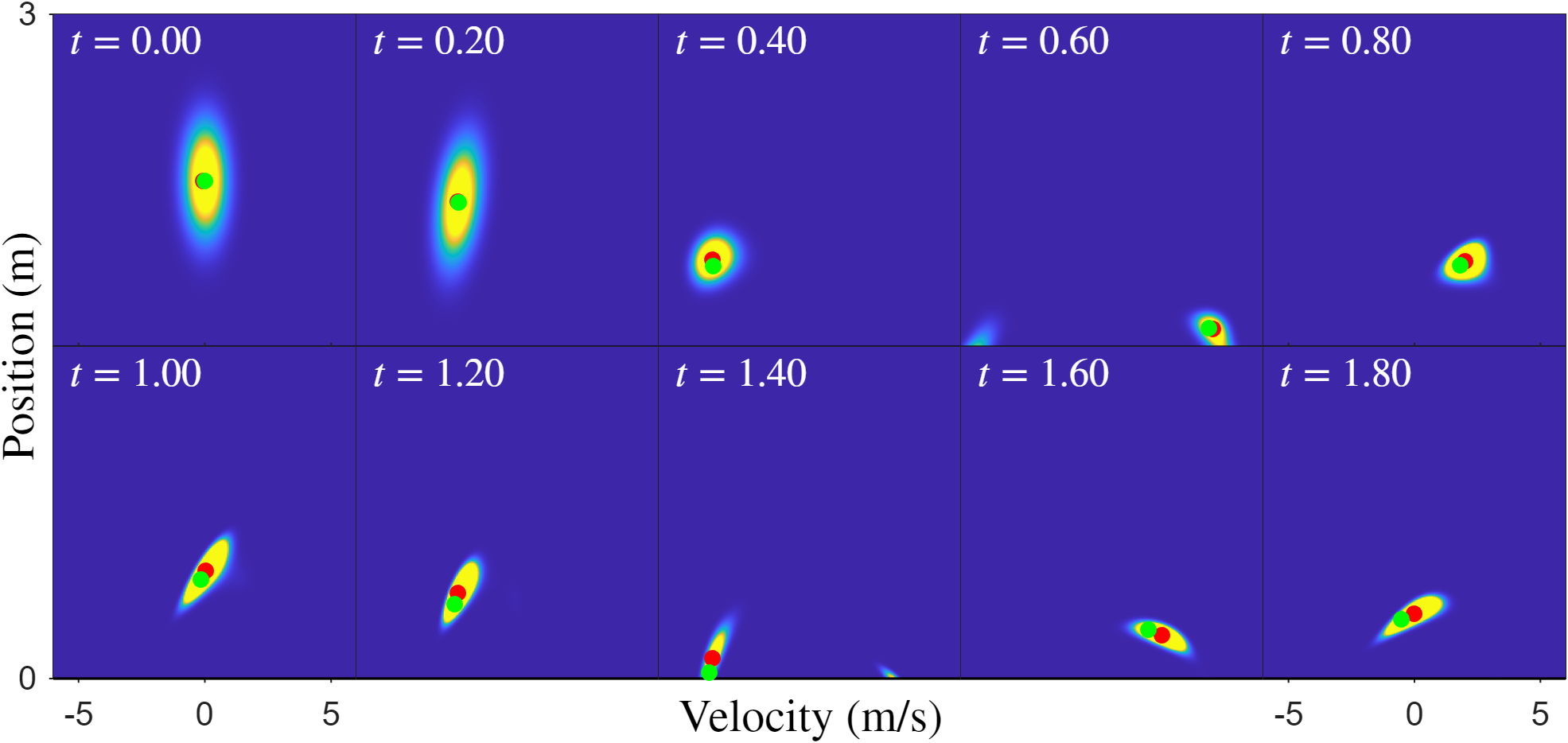}
    \caption{Posterior density snapshots for the stochastic bouncing ball system
    under intermittent measurements.}
    \label{fig:bb filtering density}
\end{figure}

\section{Conclusions \& Future Directions}
In this paper, we developed an iterative method of moment-based filtering for stochastic hybrid systems that combines finite-dimensional moment propagation with maximum-entropy reconstruction. By using Dynkin's formula together with guard-triggered reset terms, we obtained a weak evolution law that captures both continuous stochastic dynamics and discrete jump effects without directly solving the hybrid Fokker--Planck equation. In the bouncing-ball example, this framework captured impact-induced non-Gaussian structure while remaining far more tractable than direct density-based propagation.

Several interesting directions remain open. Once one leaves low-dimensional examples, the central question is not merely how to propagate uncertainty, but how to choose a finite representation rich enough to retain the essential hybrid features of the distribution. Moments together with maximum entropy provide one principled answer, but they also inherit the familiar difficulties of truncation, closure, and the growth of complexity with dimension. In more general stochastic hybrid systems, understanding how to construct accurate and stable closures remains an important open problem \cite{7039471}. On the computational side, although maximum-entropy reconstruction is systematic and convex, 
it requires solving the convex optimization problem repeatedly, which quickly becomes a burden in hybrid systems with many modes at higher orders and in higher dimensions.
Furthermore, it requires repeated evaluation of the partition function in \eqref{eq:MEM-density}. This normalization step is a familiar source of difficulty in probabilistic inference for non-normalized models \cite{Hyvrinen2005}. Finally, hybrid geometry itself introduces additional complexities through guard sets, reset maps, and boundary fluxes. Therefore, developing more scalable versions of the present framework while addressing these issues remains an important direction for future work.

\section*{Acknowledgements} This research is supported in part by
  NSF grant DMS-2103026, and AFOSR grants FA
9550-22-1-0215 and FA 9550-23-1-0400.

\bibliographystyle{IEEEtran}
\bibliography{references}

\begin{thebibliography}{10}
\providecommand{\url}[1]{#1}
\csname url@rmstyle\endcsname
\providecommand{\newblock}{\relax}
\providecommand{\bibinfo}[2]{#2}
\providecommand\BIBentrySTDinterwordspacing{\spaceskip=0pt\relax}
\providecommand\BIBentryALTinterwordstretchfactor{4}
\providecommand\BIBentryALTinterwordspacing{\spaceskip=\fontdimen2\font plus
\BIBentryALTinterwordstretchfactor\fontdimen3\font minus \fontdimen4\font\relax}
\providecommand\BIBforeignlanguage[2]{{%
\expandafter\ifx\csname l@#1\endcsname\relax
\typeout{** WARNING: IEEEtran.bst: No hyphenation pattern has been}%
\typeout{** loaded for the language `#1'. Using the pattern for}%
\typeout{** the default language instead.}%
\else
\language=\csname l@#1\endcsname
\fi
#2}}

\bibitem{WANG2020108989}
W.~Wang and T.~Lee, ``Spectral bayesian estimation for general {Stochastic Hybrid Systems},'' \emph{Automatica}, vol. 117, p. 108989, 2020.

\bibitem{Oprea2024}
M.~Oprea, A.~Shaw, R.~Huq, K.~Iwasaki, D.~Kassabova, and W.~Clark, ``A study of the long-term behavior of hybrid systems with symmetries via reduction and the {Frobenius--Perron} operator,'' \emph{SIAM Journal on Applied Dynamical Systems}, vol.~23, no.~4, pp. 2899--2938, 2024.

\bibitem{KC2025280}
T.~K.C., W.~Clark, and T.~Lee, ``Uncertainty propagation of stochastic hybrid systems: a case study for types of jump,'' \emph{IFAC-PapersOnLine}, vol.~59, no.~19, pp. 280--285, 2025, 13th {IFAC} Symposium on Nonlinear Control Systems {NOLCOS} 2025.

\bibitem{Applebaum2009}
D.~Applebaum, \emph{{L{\'e}vy} Processes and Stochastic Calculus}, 2nd~ed., ser. Cambridge Studies in Advanced Mathematics.\hskip 1em plus 0.5em minus 0.4em\relax Cambridge University Press, 2009.

\bibitem{BECT2010357}
J.~Bect, ``A unifying formulation of the {Fokker--Planck--Kolmogorov} equation for general stochastic hybrid systems,'' \emph{Nonlinear Analysis: Hybrid Systems}, vol.~4, no.~2, pp. 357--370, 2010, {IFAC} World Congress 2008.

\bibitem{MeadPapanicolaou1984}
L.~R. Mead and N.~Papanicolaou, ``Maximum entropy in the problem of moments,'' \emph{Journal of Mathematical Physics}, vol.~25, no.~8, pp. 2404--2417, 1984.

\bibitem{teng2025max}
S.~Teng, H.~Zhang, D.~Jin, A.~Jasour, R.~Vasudevan, M.~Ghaffari, and L.~Carlone, ``{Max Entropy Moment Kalman Filter} for polynomial systems with arbitrary noise,'' in \emph{The Thirty-ninth Annual Conference on Neural Information Processing Systems}, 2025.

\bibitem{Lasserre2001}
J.~B. Lasserre, ``Global optimization with polynomials and the problem of moments,'' \emph{SIAM Journal on Optimization}, vol.~11, no.~3, pp. 796--817, 2001.

\bibitem{Lasserre2015}
------, \emph{An Introduction to Polynomial and Semi-Algebraic Optimization}, ser. Cambridge Texts in Applied Mathematics.\hskip 1em plus 0.5em minus 0.4em\relax Cambridge University Press, 2015.

\bibitem{Kay1993}
S.~M. Kay, \emph{Fundamentals of Statistical Signal Processing, Volume I: Estimation Theory}.\hskip 1em plus 0.5em minus 0.4em\relax Prentice Hall, 1993.

\bibitem{7039471}
J.~Zhang, L.~DeVille, S.~Dhople, and A.~D. Dom\'{i}nguez-Garc\'{i}a, ``A maximum entropy approach to the moment closure problem for {Stochastic Hybrid Systems} at equilibrium,'' in \emph{53rd {IEEE} Conference on Decision and Control}, 2014, pp. 747--752.

\bibitem{Hyvrinen2005}
A.~Hyv{\"a}rinen, ``Estimation of non-normalized statistical models by score matching,'' \emph{J. Mach. Learn. Res.}, vol.~6, pp. 695--709, 2005.

\end{thebibliography}

\end{document}